\documentclass[twocolumn,showpacs,aps,prb,groupedaddress]{revtex4}
\usepackage{epsfig,amssymb,amsmath}
\usepackage{amsmath}
\usepackage{graphicx}
\usepackage{amssymb}

\usepackage{epsfig}
\usepackage{latexsym}
\usepackage{amssymb}
\usepackage{graphicx}

\usepackage{color}

\begin{document}

\title{Time-bin entanglement of quasi-particles in semiconductor devices}

\author{Luca Chirolli, Vittorio Giovannetti, Rosario Fazio}
\affiliation{National Enterprise for nanoScience and nanoTechnology (NEST), Scuola Normale Superiore and Istituto Nanoscienze-CNR,  piazza dei Cavalieri 7, I-56126 Pisa, Italy}
\author{Valerio Scarani} 
\affiliation{Centre for Quantum Technologies \& Department of Physics, National University of Singapore, 3 Science Drive 2, Singapore 117543}


\begin{abstract}
A scheme to produce time-bin entangled pairs of electrons and holes is proposed. It is based on a high frequency time-resolved single-electron source from a quantum dot coupled to one-dimensional chiral channels. Operating the device in the weak tunneling regime, we show that at the lowest order in the 
tunneling rate, an electron-hole pair is emitted in a coherent superposition state of different time bins determined by the driving pulse sequence.
\end{abstract}

\maketitle

\section{Introduction} 

Because of the very long coherence times, electronic states at the edge of a two-dimensional electron 
gas in the integer quantum Hall effect (IQHE) regime~\cite{DATTA,BEEN} are ideal systems for designing 
of coherent electronics circuitry or to implement quantum information processing.
In particular, electronic versions of several optical interferometers have been realized~\cite{INTERF,NederNature07,NederPRL06,NederPRL07,RoulleauPRL08,RoulleauPRBr07,RoulleauPRL09,
LitvinPRB07,BieriPRB09} using continuous electron sources. Recently it was also shown that high frequency gate  modulation  can realize single electron sources (SESs)~\cite{GlattliScience07,MahePRB,Blumenthal}, that allow one  to inject in a  controlled and coherent way single electrons and holes onto an edge state. 
Exploiting the beam splitting technique via quantum point contact (QPC), the electronic  wave packet produced by the SES could also be split, recombined, or  coherently guided toward different paths, via the application 
of external gate voltages~\cite{JAN1,JAN}. Altogether, SESs, QPCs, and electronic waveguides represent 
the necessary toolbox toward the realization of an electronic version of the numerous protocols developed 
in quantum optics~\cite{Tittel}. The only qualitative difference is in the measurement process, in that 
photons are usually absorbed by photodetectors and are studied in terms of $n$-photon coincidence correlation functions, whereas electrons are characterized by currents and higher-order momenta. 
In Ref.~[\onlinecite{JAN}] this formal equivalence was used to characterize two-particle non-local effects originating via collision and proper post-selection from two independent  SESs whose output states were coherently mixed at a~QPC.  

A full exploitation of the capabilities of quantum information however requires the production and 
manipulation of entangled states. Several schemes to generate  entangled states in multi-terminal 
mesoscopic conductors have been proposed so far (see~[\onlinecite{BEEN}] and references therein). 
The realization of the SES~\cite{GlattliScience07} opens possibilities to realize entangled states.
Here we exploit the coherence of time-resolved single-electron wave packets  at the output of a single SES, operated in the weak tunneling regime to generate superpositions of an electron-hole ({\it e}-{\it h}) pair produced at different times. Differently from Ref.~[\onlinecite{JAN}] our scheme does not rely on collisional  mechanisms followed by post-selection: As a consequence, in our case the {\it e}-{\it h} pair emerges from 
the device in an entangled state without the need of any filtering processes. The resulting output closely reminds us of the bi-photon state produced by a down-conversion nonlinear crystal 
(see e.g. Ref.~[\onlinecite{beam}]) or, more precisely, the entangled photon holes states~\cite{Beenakker03,FRANSON1} produced via two-photon absorbing processes. It admits a representation 
in terms of time-bin entanglement~\cite{TIMBIN,ScaraniPRL04} whose two-particle correlations we characterize by performing current cross-correlation measurements at the output of a Franson 
interferometer~\cite{Franson} (the latter being an interferometric setting which is specifically designed 
to detect coherence properties of sources emitting pairs of correlated particles).

The paper is organized as follows.
In Sec.~\ref{sec:model} we present the setup and describe in brief the basic idea of the proposal. 
The dynamics of the device is discussed in Sec.~\ref{sec:dyn} and in Sec.~\ref{sec:nonidealcase} we discuss deviation from the ideal case. Conclusions and remarks are given in Sec.~\ref{sec:conclusion}.

\begin{figure}[t]
 \begin{center}
  \includegraphics[width=8cm]{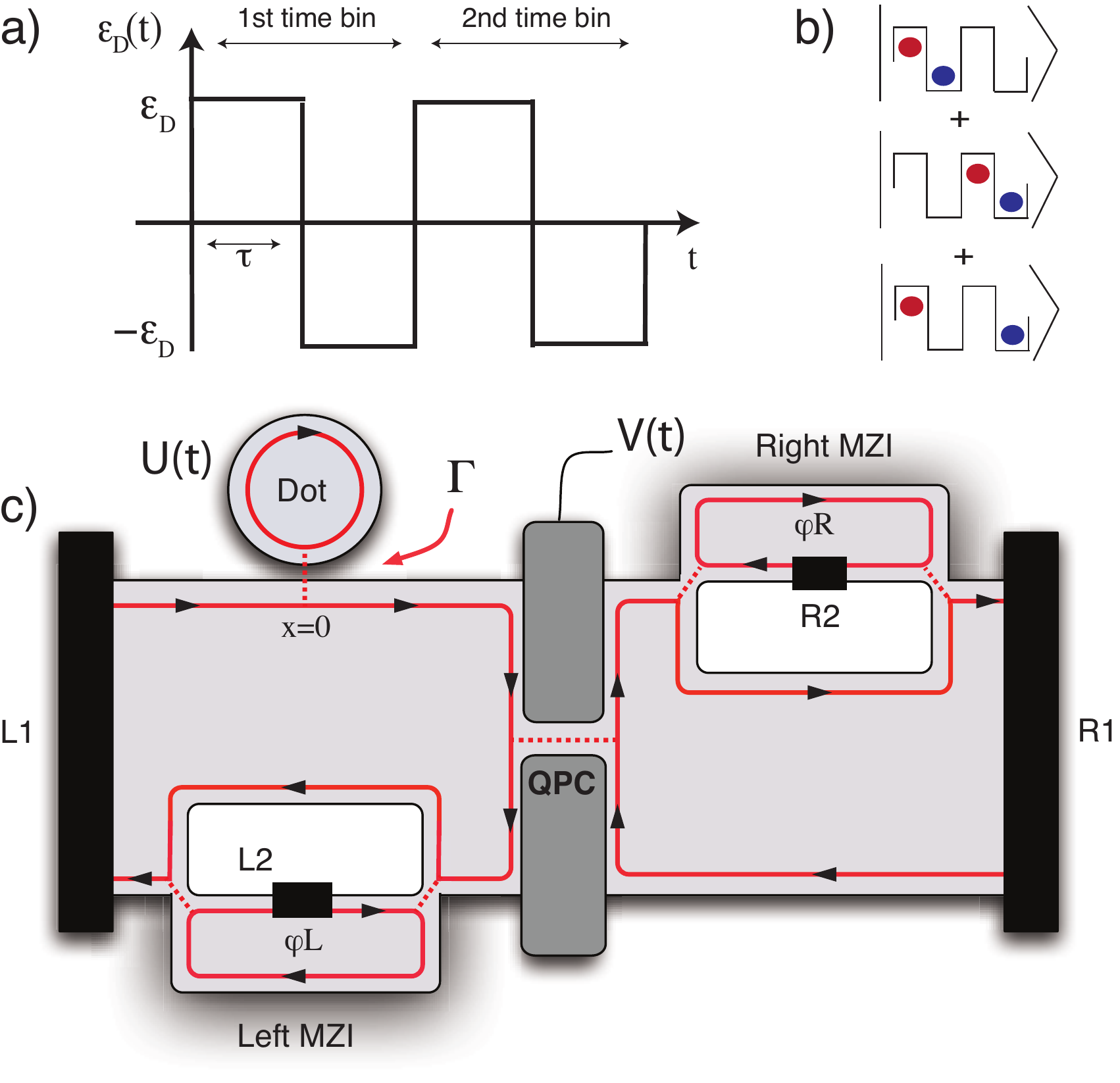}
    \caption{(Color online) (a) Driving sequence of the dot level showing the first and the second time-bin. (b) Pictorial representation of the state of the system at first order in the tunneling rate $\Gamma$. Here the red (gray) and blue (dark gray) dots represent, respectively, the electron and hole emitted from the SES, while the slots inside the brackets define the various time bins of the sequence. (c) Implementation of a Franson interferometer for an IQHE architecture at $\nu=1$. A quantum dot driven by a gate potential $U(t)$ generates an {\it e}-{\it h} pair on the edge channel at position $x=0$. The electron and hole propagate freeley until they reach the QPC  driven by the time-dependent voltage $V(t)$ that splits them and send the electron toward the right-hand MZI~\cite{INTERF} and the hole toward the left-hand one. The black elements are ohmic contacts to different reservoirs, all kept grounded.\label{Fig1}}
 \end{center}
\end{figure}

\section{The setup} 
\label{sec:model}

We start with an intuitive description of the model whose validity will be checked later. As shown in 
Fig.~\ref{Fig1} (c), the SES is described as a quantum dot whose energy levels are externally controlled 
via a time-dependent  voltage gate $U(t)$, and which is connected to an  edge state of the IQHE effect 
at a filling factor $\nu=1$ via a QPC characterized by a tunneling amplitude~$A$ which  can be externally modulated~\cite{GlattliScience07,MahePRB}. In this setup the linear dispersion relation of the edge state 
close to the Fermi energy gives rise to a constant velocity of propagation $v_F$ for the electronic wave packets: Consequently, a particle generated by the SES at the position $x=0$ of the edge will be found translated by an amount $v_Ft$ after time $t$ has elapsed. As in Refs.~[\onlinecite{albert} and [\onlinecite{MahePRB}], we consider then an elementary driving sequence of duration $2\tau$ consisting of two  subsequent movements in which  we first rapidly rise the dot energy level  above the Fermi sea and keep it there for a time $\tau$,  and then we rapidly lower it  below the Fermi sea and keep it there for a time $\tau$. 
Assuming  $\tau$ to be much smaller than the dot escaping time $1/\Gamma$ ($\Gamma\propto A^2$), 
a pair of well-separated time-resolved electron and hole can be created on the edge channel with probability $(2\Gamma\tau)^2\ll 1$, the electron being localized in the first half of the time-bin and the hole being 
localized in the second half  (other  processes  in which no excitation or just a single excitation is  emitted 
can be neglected  since they do not contribute to the current cross-correlation measurements). This  weak 
tunneling regime is crucial for the results we present and is actually opposite to the one usually considered 
in other proposals~\cite{JAN1,JAN} (to stress this fact from now on we will refer to our source as a weak tunneling SES, or wt-SES in brief). Still it is well within the reach of current experimental capabilities: For instance, considering that $1/\Gamma$  of the order of  $10$ ns can be achieved while preserving
the coherence of the process~\cite{GlattliScience07,MahePRB}, the rising time of the dot energy level can 
be set to be of order of few  ps, while $\tau$ can be taken to be of the order of a fraction of ns. Notice also 
that during the driving we never leave the dot at resonance with the Fermi level for an extended period of 
time:  As experimentally verified in Ref.~[\onlinecite{MahePRB} and [\onlinecite{albert}], this allows us to 
avoid the collateral  generation of pairs induced by resonance effect between the Fermi level and the 
dot~\cite{Levitov08}.

Consider hence the case in which the dot is initially charged and the elementary driving sequence is 
repeated several times, say, twice, as depicted in  Fig.~\ref{Fig1}(a). In the weak tunneling regime the 
global state of the system [dot + one-dimensional (1D) line] at the lowest order in the tunneling amplitude 
$A$ contains three contributions. First, we have an unperturbed component in which the dot is still charged and no excitations are produced in the 1D channel (zero-order contribution in  $A$). Then there is a contribution proportional to  $A$ in which the dot electron has emerged from the dot but no holes have 
been subsequently produced by the driving sequence. Last, we have a  term proportional to $A^2$ 
describing the case in which no charge is trapped in the dot and the 1D line contains a delocalized 
{\it e}-{\it h} pair: As anticipated, this is the only component of the state which can contribute to the current cross correlations we perform at the output of the setup, the probability of the event being proportional to 
$|A|^4$ -- see below. If the driving process is  kept coherent, it is described as a coherent superposition of 
an {\it e}-{\it h} pair emitted in the first time bin, an {\it e}-{\it h} pair emitted in the second time bin and an 
{\it e}-{\it h} pair with the electron emitted in the first time bin and a hole emitted in the second time bin, 
-- see Fig.~\ref{Fig1}(b). We can hence represent it as the following vector:
\begin{equation}
\label{Eq:ehTimeBin}
|\Psi\rangle\,\propto\,
|1,0\rangle_e|1,0\rangle_h+|0,1\rangle_e|0,1\rangle_h+|1,0\rangle_e|0,1\rangle_h\;,
\end{equation}
where $|1,0\rangle_{e,h}$ is  the state with one electron (hole) in the first time bin (first label in the ket) 
and none in the second time bin (second label in the ket), while $|0,1\rangle_{e,h}$ is the state with no 
electron (hole) in the first time bin and one electron (hole) in the second time bin. We stress that 
$|\Psi\rangle$ is only the second order contribution in the tunneling amplitude  $A$ to the real full state 
of the system, which we can use to evaluate the current cross correlations at the output of the device 
(in other words, it is the component of the full state that gets post-selected by our measuring apparatus).

Equation~(\ref{Eq:ehTimeBin}) represents  an  entangled state of two qubits: For instance, it violates the Clauser-Horne-Shimony-Holt (CHSH) inequality \cite{Bib:CHSH} up to 
\begin{eqnarray}
\mathrm{CHSH}(\Psi)\,=\,2\sqrt{13}/3\,\approx\,2.404\;,\label{CHSH}
\end{eqnarray} for suitable measurements~\cite{HorodeckiPLA95}. Specifically, identifying  $|1,0\rangle$ 
and $|0,1\rangle$ with the eigenstates of $\sigma_z$,  the value in (\ref{CHSH}) is achieved by using the following set of local observables
\begin{eqnarray} 
\Sigma_e^{(0)}&=&(\sigma_z+2\sigma_x)/\sqrt{5}\;,
\nonumber \\
\Sigma_e^{(1)}&=&(\sigma_x-2\sigma_z)/\sqrt{5}\;,  \label{SIGMAe}
\end{eqnarray}
 for the electronic part of Eq.~(\ref{Eq:ehTimeBin}), and 
\begin{eqnarray}
\Sigma_h^{(0)}&=&(4\sigma_x-7\sigma_z)/\sqrt{65}\;,\nonumber \\
\Sigma_h^{(1)}&=&(\sigma_z+8\sigma_x)/\sqrt{65}\;, \label{SIGMAh}
\end{eqnarray} 
for the hole part.

The time-bin entanglement of the state in Eq.~(\ref{Eq:ehTimeBin}) can be detected by means of a Franson interferometer~\cite{Franson}. The two sub-systems (here, the electron and the hole) must be separated and sent to unbalanced Mach-Zehnder interferometers (MZIs), at the output of which coincidences are recorded.  Furthermore, on each side, single-qubit measurements can be implemented probabilistically by adapting the phase delays and the transmittivity of the second beam-splitter (BS) of the MZI. Therefore, in principle, in the Franson setup the value of CHSH or of any other entanglement witness can be measured. This is well known and we refer the reader to the optical implementations for details \cite{optics}. Here, we rather have to discuss how a Franson setup can be realized for our {\it e}-{\it h} pairs generator. The electron and the hole in each time bin are separated by means of a time-dependent QPC that acts as a switch, sending the electron and 
the hole, respectively, toward different MZIs, which are implemented along the lines of Ref.~[\onlinecite{INTERF}], as depicted in Fig.~\ref{Fig1}. This separation is a challenging task: The potential $V(t)$ of the 
QPC has to rise on a time scale equal or smaller to the inverse frequency of the gate potential $U(t)$ on the dot, in such a way that the switch perfectly ``cuts" the wave function. The difference $\ell$ between the long and the short path in each MZI is chosen equal to $2v_F \tau$. On the one hand, this guarantees that no single-particle interference arises within the MZIs which then operate as effective, probabilistic (but coherent) delay lines. On the other hand, this allows us to align the time slot associated with the long path of a particle belonging to the first time bin with the time slot of the short path associated with second time bin. The current cross correlations are finally measured  at the outputs of the setup, i.e.,
\begin{eqnarray}
\delta{\cal C}_{R_j,L_i}=\langle\delta I_{R_j}(t')\delta I_{L_i}(t)\rangle\;,
\end{eqnarray} 
with  $\delta I_{\alpha_i}(t)= I_{\alpha_i}(t)-\langle I_{\alpha_i}(t)\rangle$, where for $\alpha=L,R$ and $i,j=1,2$, $I_{\alpha_i}(t)$ is the current operator at the $\alpha_i$ port evaluated at time $t$, and where 
$\langle \cdots \rangle$ stands for the expectation value on the initial state of the system (notice 
that different times $t\neq t'$ have to be chosen since the electron and hole are time shifted within the 
same time bin). This allows to post-select the events we are interested in, discarding single-particle currents. 

The transmittivities of the BSs can be engineered as in Ref.~[\onlinecite{INTERF}] and do not require special discussion. For the sake of simplicity, we first assume to fix them at $50\%$ as in the original proposal by Franson~\cite{Franson}: Indeed, this choice is sufficient to detect a signal which is sensitive to the superposition of Eq.~(\ref{Eq:ehTimeBin}) [instead to recover $\mathrm{CHSH}(\Psi)$ one would 
need to adjust the transmittivities as detailed at the end of the section].  In order to have a phase difference between the two arms of a MZI, we invoke instead the Aharonov-Bohm (AB) effect. Accordingly, the statistics 
of the coincidence events will be sensitive to a non-local two-particle AB phase~\cite{ButtikerPRL04} that 
gives rise to interference fringes that can witness entanglement in the {\it e}-{\it h} pair. To see this, consider that  each of the three input terms of Eq.~(\ref{Eq:ehTimeBin}) is mapped after the MZIs into the sum of four 
states. Out of the 12 total contributions, three will be indistinguishable when revealing the proper coincidence counts:  The event where the {\it e}-{\it h} pair is generated in the first time-bin and both 
particles choose the long paths in the MZIs, i.e.,
\begin{eqnarray}
|1,0\rangle_e |1,0\rangle_h \rightarrow e^{i\varphi_R-i\varphi_L}|0,1\rangle_e|0,1\rangle_h+\cdots\;,
\end{eqnarray} 
(here $\varphi_{R,L}$ are the AB phases computed  with a gauge for which only particles traveling along the long arms of the MZIs acquire a phase, the  difference in sign arising from the opposite charge for electron and the hole);  the event where the {\it e}-{\it h} pair is generated in the second time bin and both the particles choose the short paths of the MZIs, i.e.,  
\begin{eqnarray}
|0,1\rangle_e|0,1\rangle_h\rightarrow |0,1\rangle_e|0,1\rangle_h + \cdots\;;
 \end{eqnarray}
finally, the event where the electron is generated  in the first time bin, the hole in the second time bin and choose, respectively, the long and the short paths, i.e., 
 \begin{eqnarray}
 |1,0\rangle_e|0,1\rangle_h\rightarrow  e^{i\varphi_R} |0,1\rangle_e|0,1\rangle_h + \cdots\;.
 \end{eqnarray}
Measuring then coincidence events in the second time-bin at the output of the MZIs (i.e., $|0,1\rangle_e|0,1\rangle_h$) will then produce a signal which is sensitive to the coherent superposition of Eq.~(\ref{Eq:ehTimeBin}), i.e., 
 \begin{eqnarray}
\delta{\cal C}_{R1,L1}&=&-({\Gamma}/{4})^2\; |1+e^{i\varphi_L}+e^{i\varphi_L-i\varphi_R} |^2\;. \label{COINC}
\end{eqnarray}
The three terms correspond to the three vectors which compose Eq.~(\ref{Eq:ehTimeBin})  and display coherent oscillations. In particular, a dependence on a non-local two-particle AB phase appears as the difference of the AB phases of the left- and right-hand MZIs.
 
 \subsection{Measuring  CHSH correlations}
 \label{sec:CHSH}
 
We analyze in more details how to use our Franson setting to  measure the value of CHSH$(\Psi)$. We 
remind that to do so we need to perform local measurements on the two subsystems (electron and hole separately). As anticipated, this can be done in a probabilistic fashion by properly adjusting the transmittivities of the BS of the setup. Indeed, let us consider  the effect of an asymmetric MZI, characterized by transmission $\sqrt{s}$ and reflection $i\sqrt{r}$ amplitudes of the second BS, while keeping the first BS symmetric. Let us suppose also that the generic state of, say, the electron before the right-hand MZI can be written as $|\phi\rangle_{\rm in}=\alpha|1,0\rangle_e+\beta|0,1\rangle_e$. By focusing on the second time bin at the outputs of the MZI, the electron will come out with an amplitude 
\begin{eqnarray}
S_1=-(\alpha e^{i\varphi_R}\sqrt{r}-\beta\sqrt{s})/\sqrt{2}\;,
\end{eqnarray}  from the output 1 and with amplitude 
\begin{eqnarray}
S_{2}=i(\alpha e^{i\varphi_R}\sqrt{s}+\beta\sqrt{r})/\sqrt{2}\;,
\end{eqnarray} 
from output 2. Defining the rotated input qubit states 
\begin{eqnarray}
|u_+\rangle&=&e^{-i\varphi_R}\sqrt{s}|1,0\rangle_e+\sqrt{r}|0,1\rangle_e\nonumber \\
|u_-\rangle&=&e^{-i\varphi_R}\sqrt{r}|1,0\rangle_e-\sqrt{s}|0,1\rangle_e\end{eqnarray} 
it follows that  $\langle u_+|\phi\rangle_{\rm in}=-i\sqrt{2}S_2$ and $\langle u_-|\phi\rangle_{\rm in}=-\sqrt{2}S_1$, and it becomes clear that in the second time-bin outputs one reads the results of the measurements of $|\phi\rangle_{\rm in}$ on the eigenstates of 
\begin{eqnarray}
(s-r)\sigma_z+2\sqrt{sr}(\cos(\varphi_R)\sigma_x+\sin(\varphi_R)\sigma_y)\;,
\end{eqnarray}
where the Pauli matrices are written in the basis $|1,0\rangle_e$ and $|0,1\rangle_e$. The probability of  success is $1/2$, provided that $|\alpha|^2+|\beta|^2=1$. According to this analysis we can hence realize 
the  observables~(\ref{SIGMAe}) by taking $\varphi_R=0$ and $s_{0}=(1+1/\sqrt{5})/2$, $s_{1}=(1-2/\sqrt{5})/2$ for the MZI on the right-hand side of Fig.~\ref{Fig1}, and the observables~(\ref{SIGMAh}) by taking instead 
$\varphi_L=0$ and  $s_{0}=(1-7/\sqrt{65})/2$, and $s_{1}=(1+1/\sqrt{65})/2$ for the MZI on the left-hand side.

\section{Microscopic model} 
\label{sec:dyn} 
 
We now turn to a more quantitative model  where the dynamics of the wt-SES  is described via a time-dependent Hamiltonian~\cite{Levitov08}  of the form
\begin{eqnarray}
H(t)=\varepsilon_D(t)d^{\dag}d+\sum_{k}\varepsilon_kc^{\dag}_{k}c_{k} + H_{\rm tun}(t)\;,
\end{eqnarray} 
where $d$ and $c_k$ are, respectively, the Fermionic annihilation operators of the dot energy level and of 
the 1-D free-electron modes associated with the chiral IQHE  edge channel at $\nu=1$, while 
$H_{\rm tun}(t)$ is the tunneling term. We take the dot energy $\varepsilon_D(t)$ as in Fig.~\ref{Fig1}(a) 
while, assuming linear dispersion around the Fermi energy (set to zero), we write the energy levels of the $c_k$ modes as $\varepsilon_k=\hbar v_F k$ (nonlinear corrections being typically negligible in IQHE systems for small bias voltages). In the weak-coupling regime we consider a (time dependent) tunneling  amplitude peaked around the resonance value $k_D(t) =\varepsilon_D(t)/\hbar v_F$ associated with the instantaneous dot energy~\cite{MahePRB} within a bandwidth $BW=\hbar v_F\Delta k$, i.e., 
\begin{eqnarray}
H_{\rm tun}(t) =A\sum_{k\in[k_D(t) ,\Delta k]}(d^{\dag}c_k+c^{\dag}_kd)
\;, 
\end{eqnarray}
and assume $k_D>\Delta k$. The spread $\Delta k$ is associated with the uncertainty in the emission 
position of the electron on the 1D channel, taking  into account  the not perfect point-like coupling between 
the dot and the edge, allowing hence the tunneling hopping to extend over a range 
$\Delta x\simeq 2\pi /\Delta k$.  A natural bound on $\Delta x$ can be set as  $\Delta x\leq R_D$, where 
$R_D$ is the linear dimension of the quantum dot. The latter, however, is directly related with the dot energy-level spacing $\Delta=2 \varepsilon_D$ via the expression $\Delta\simeq \hbar^2(2\pi/R_D)^2/2m^*$, $m^*$ being the effective mass of the electron ($m^*=0.068~m_0$ for GaAs heterostructures). It follows a minimum bandwidth on order 
 \begin{eqnarray}
 BW_0=2\pi \hbar v_F/R_D=\sqrt{2m^*v_F^2\Delta}\;.\end{eqnarray}  To put some number we notice in the experiment of Ref.~[\onlinecite{MahePRB}]  one has $\Delta\simeq 4.4~{\rm K}$ ($\Delta\simeq 0.36~{\rm meV}$). Choosing $v_F\simeq 10^{4}~{\rm m/s}$, we get $BW_{0}\simeq~0.2~{\rm meV}$. Therefore, we 
see that the validity of the condition $k_D>\Delta k$, ensured if $BW_0/\Delta\leq BW/\Delta<1$, can be satisfied. 

In the interaction picture with respect to free evolution of the system, the Hamiltonian becomes hence 
\begin{eqnarray}
H_I=A\sqrt{L_c}\,e^{i\varphi_D(t)}\,d^{\dag}\,\psi_{k_D(t)}(-v_Ft)+{\rm h. c.}\;,
\end{eqnarray}
where 
\begin{eqnarray}
\varphi_D(t)=\frac{1}{\hbar}\int^tdt'\varepsilon_D(t')\;, \end{eqnarray}
is the dot dynamical phase, $L_c$ is the channel length, and 
\begin{eqnarray}
\psi_{k_D}(x)=\frac{1}{\sqrt{L_c}}\sum_{k\in[k_D,\Delta k]}e^{ikx}c_k\;,
\end{eqnarray}
is the field operator in position space of an electron propagating with mean momentum $k_D$. When applied to the Fermi sea, the operator $\psi^{\dag}_{k_D}(x)$ adds an electron above the Fermi sea only if $k_D>0$ 
($k_F=0$). If the initial dot energy is negative, $\varepsilon_D<0$, the electronic field operator $\psi_{-|k_D|}(x)$, when applied to the Fermi sea, creates a hole of average momentum $|k_D|$.  Field operators at positions $x$ and $x'$ satisfy canonical anti-commutation rules if $|x-x'|\gg \Delta x$, which amounts to a coarse graining of the position resolution. Due to the linearity of the channel dispersion, we have a one-to-one mapping between position and time, which implies a coarse graining in the time coordinate: Fields at times $t$ and $t+\delta t$, with $\delta t\lesssim 2\pi/(v_F\Delta k)$, are indistinguishable and the anti-commutation relations at times differing by $\delta t$ have to be understood as at equal time. 

Consider now the time evolution of the input state $|1\rangle_D|\Theta_F\rangle$, where $|1\rangle_D$ describe the charged state of the dot and  $|\Theta_F\rangle$ is  the free Fermi-sea state of the edge. In the interaction picture the first  correction that gives rise to nonzero current cross correlations at the output is described by the vector
\begin{widetext}  
\begin{eqnarray}
\int_0^t dt'\int_0^{t'} dt''e^{i\varphi_D(t',t'')}\psi_{k_D(t')}(-v_Ft')\psi^{\dag}_{k_D(t'')}(-v_Ft'')
|\Theta_F\rangle\;,
\end{eqnarray}
\end{widetext}
which exhibits an electron of momentum $k_D(t'')$ at time $t''$ and a hole with momentum $-k_D(t')$ at 
time $t'>t''$.  Assuming piece-wise constant $k_D$, sharp transitions, and neglecting transient effects,
we can set $k_D>0$ for $0<t<\tau$ and $k_D<0$ for $\tau<t<2\tau$, as shown in Fig.~\ref{Fig1}(a). Since 
the two consecutive time slots do not overlap, one can create a state with the electron localized within 
$\ell/2<x<\ell$ and a hole localized within $0<x< \ell/2$ (here $\ell\equiv 2v_F\tau$). Defining thus the 
electron and hole field operators $\psi_e(x)\equiv\psi_{k_D}(x)$, $\psi_h^{\dag}(x)\equiv\psi_{-k_D}(x)$, 
we can write the above vector after time $t=4\tau$ as 
\begin{equation}\label{Eq:eh-state}
|\Psi\rangle=-\frac{\Gamma}{v_F}\int dx_1dx_2\phi^{(2)}(x_1,x_2)\psi^{\dag}_h(x_1)\psi_e^{\dag}(x_2)|\Theta_F\rangle, 
\end{equation}
with the escaping rate $\Gamma=A^2L_c/\hbar^2v_F$ and the {\it e}-{\it h} wave function 
\begin{eqnarray}
\phi^{(2)}(x_1,x_2)&=&{\rm Step}(x_1,x_2)+e^{-ik_D\ell}{\rm Step}(x_1+\ell,x_2)\nonumber \\
&&+e^{-ik_D\ell}{\rm Step}(x_1+\ell,x_2+\ell)\;,
\end{eqnarray}
expressed in terms of the Heaviside distribution $\theta(x)$ through the identity
 \begin{eqnarray}
 {\rm Step}(x_1,x_2)&=&e^{ik_D(x_1+x_2)}\theta(x_1+\ell)\theta(-x_1-\ell/2)\nonumber \\
& &\times \theta(-x_2)\theta(x_2+\ell/2)\;.\end{eqnarray} 
This is the  second quantization representation of the state of~Eq.~(\ref{Eq:ehTimeBin}) which defines, 
in the interaction picture, the electron-hole distribution on the 1-D channel that connects the dot with the 
time-dependent QCP. The propagation through the latter can then be described in the Landauer-B\"uttiker formalism \cite{LandauerButtiker} by introducing a time-dependent transmission amplitude toward the 
right-hand MZI, $s_R(t)$, and reflection amplitude toward the left-hand MZI, $s_L(t)$, with $|s_L|^2+|s_R|^2=1$,  ${\rm Re}~s^*_Ls_R=0$. We assume sharp transitions and neglect transient effects, such that the QPC is totally transmitting during the first half of the time bin and it is totally reflecting during the second 
half of the time bin (i.e., $|s_R(t)s_L(t+\tau)|=1$). For $\alpha = R,L$ and $j=1,2$, the field at the $\alpha_j$ output of the setup can then be expressed as 
\begin{eqnarray}
\psi_{\alpha_j}(x,t)&=&i^{j-1}s_{\alpha}(t)\left[\psi(x-v_Ft)\right.\nonumber\\
&+&\left.(-)^{j-1}e^{i\varphi_{\alpha}}\psi(x+\ell- v_F t)\right]/2\;, 
\end{eqnarray}
where $\varphi_{\alpha}$ the AB phases and where  $\psi(x-v_Ft)$ is the drifting full free field operator 
\begin{eqnarray}
\psi(x)=\frac{1}{\sqrt{L_c}} \sum_k e^{ikx} c_k\;,
\end{eqnarray}
of the channel that  connects the dot with the time-dependent QCP which, neglecting the Fermi-sea contribution, in our case can be approximated as    
 $\psi(x)\approx\psi_e(x)+\psi^{\dag}_h(x)$. 
The correlation $\delta C_{R_1,L_1}$  of Eq.~(\ref{COINC}) is then computed by observing that the 
current density operator  at position $x$ in the 1D channel associated with the output port $\alpha_j$ 
can be expressed as 
\begin{eqnarray}
{I}_{\alpha_j} (x,t)=v_F{\psi}_{\alpha_j}^{\dag}(x,t){\psi}_{\alpha_j}(x,t)\;.
\end{eqnarray}
Setting $x_e=x-v_Ft$, $y_h=x_e-v_F\tau$ to compensate for the time shift between the electron and 
the hole, the resulting expression for the cross-correlator results in 
\begin{eqnarray}
\delta C_{R_1,L_1}= -\frac{v_F^2}{16} \Big| \sum_{a,b=0,1}  {\cal G}(x_e+a \ell, y_h + b  \ell) e^{i a \varphi_R - i b \varphi_L}\Big|^2,\nonumber\\
\end{eqnarray}
in terms of the Green's function ${\cal G}(x,y)\equiv\langle\psi^{\dag}(x,t)\psi(y,t)\rangle$, which at the lowest order in $\Gamma$ is
 \begin{eqnarray}
 {\cal G}(x_e, y_h) &=&\langle\Theta_F| \psi^{\dag}(x_e)\psi(y_h)|\Psi\rangle\nonumber\\
 &+&\langle\Psi| \psi^{\dag}(x_e)\psi(y_h)|\Theta_F\rangle\;.
 \end{eqnarray}
Choosing then $k_D\ell=2\pi$ and measuring the currents at the central peak $-3/2<x_e/\ell<-1$ 
one finally gets~Eq.~(\ref{COINC}).

\section{Non ideal case}
\label{sec:nonidealcase}

Until now we have considered an ideal situation of a linear dispersion channel, a zero-temperature working regime, and no dephasing process has been taken into account. Quadratic deviation from a linear dispersion implies that each $k$ state propagates at his own speed, yielding a spread of the wave function. Typically, electron-like states with energy above the Fermi energy propagate slightly faster than hole-like states with energy below the Fermi energy. Consequently, the time bin becomes smeared and adjacent time bins develop an overlap as the electron-hole pair propagates along the channel, making the synchronization of the electron and hole more difficult.

After the state Eq.~(\ref{Eq:eh-state}) has been produced at position $x=0$, it propagates into the channel and interaction with the environment reduces the degree of time-bin entanglement of the electron-hole pair. Typically, in the IQHE the single-particle coherence length has been proved to be very long, on the scale of hundreds of micrometers. For dc-biased MZI interferometers a direct measurement of the single-particle coherence length has been reported in Ref.~[\onlinecite{RoulleauPRL08}] by monitoring the decrease of the visibility of single-particle Aharonov-Bohm oscillation in MZIs of different sizes. An observed $1/T$ dependence versus the temperature has been attributed to thermal noise of the dissipative part of the finite frequency coupling impedance between the environment and the reservoirs. Short- and long-range interactions as well as curvature of the fermion dispersion have been ruled out due to an expected different dependence on temperature. As far as two-particle  processes are concerned, an experimental measurement of two-particle Aharonov-Bohm phase  has been performed by Neder {\it et al.} [\onlinecite{NederNature07}], where a visibility on order of $70\%$ for a dc-biased case has been reported. 

In the case that we consider, the electronic reservoirs are kept at the same bias and the non-equilibrium nature of the excitation is entirely due to the dot driving. At the moment no experimental test of single- or two-particle interferometry with SESs or wt-SESs has been reported to our knowledge. On the other hand, a theoretical study has shown that, for a single-electron wave packet injected in chiral Luttinger liquid, the deformation of the wavepacket due to electron-electron interaction can be partly undone by a suitable voltage pulse [\onlinecite{LebedevPRL11}], whereas in a real device a capacitive Coulomb interaction may add dissipation into the system~[\onlinecite{DegiovanniPRB09}].

\section{Conclusions} 
\label{sec:conclusion} 

By operating a SES in the weak tunneling regime, we have proposed a scheme to generate a time-bin entangled state of an {\it e}-{\it h} pair which can be detected via current cross-correlation measurements 
at the output of a Franson interferometer. Within the range of validity of our approximations the only higher-order excitations produced in a two time-bin cycle are $|1,1\rangle_e |1,0\rangle_h$ and $|1,1\rangle_e|1,1\rangle_h$, which are clearly discriminable since they bring about a different charge (hence, by 
monitoring all the four outputs of the Franson interferometer, it is possible, in principle, to discard their contribution). Transient effects, together with all higher-order terms, also contribute to the shape of the electron-hole wave function, which within our first-order approximation only amounts to a phase, resulting in a delocalization of the electron and hole in the two time bins and a degradation of the signal.
\newline

We acknowledge financial support by the FIRB-IDEAS project, RBID08B3FM, EU Project IP-SOLID, 
and by the National Research Foundation and the Ministry of Education of Singapore.

\end{document}